\documentclass[11pt]{article}

\usepackage{amsmath}
\usepackage{graphicx,psfrag,epsf}
\usepackage{enumerate}
\usepackage{natbib}

\usepackage[footnotesize]{caption}
\usepackage{wrapfig}
\usepackage{setspace,verbatim}

\usepackage{graphicx,natbib}
\usepackage{subcaption}
\usepackage{float}
\usepackage{wrapfig}

\usepackage{amsthm,amsmath,color}
\usepackage{amssymb}
\usepackage{latexsym}
\usepackage{float}
\usepackage{amsfonts}
\usepackage{bbm}
\usepackage{longtable}
\usepackage{booktabs}
\usepackage{multirow}

\newcommand*{\defeq}{\mathrel{\vcenter{\baselineskip0.5ex \lineskiplimit0pt
                     \hbox{\scriptsize.}\hbox{\scriptsize.}}}%
                     =}

\newtheorem{theorem}{Theorem}

\newtheorem{remark}{Remark}

\begin{document}

\nocite{*}

\pagestyle{empty}

%%%%%%%%%%%%%%%%%%%%%%%%%%%%%%%%%%%%%%%%%%%%%%%%%%%%%%%%%%%%%%%%%%%

\begin{center}
\begin{Large}{\bf An exposition of the false confidence theorem \\}
\vspace{.25in}
\end{Large}

Iain Carmichael \ and \ Jonathan P Williams \\[.15in]
University of North Carolina at Chapel Hill
 
\begin{abstract}
A recent paper presents the ``false confidence theorem''  (FCT) which has potentially broad implications for statistical inference using Bayesian posterior uncertainty. This theorem says that with arbitrarily large (sampling/frequentist) probability, there exists a set which does \textit{not} contain the true parameter value, but which has arbitrarily large posterior probability. Since the use of Bayesian methods has become increasingly popular in applications of science, engineering, and business, it is critically important to understand when Bayesian procedures lead to problematic statistical inferences or interpretations. In this paper, we consider a number of examples demonstrating the paradoxical nature of false confidence to begin to understand the contexts in which the FCT does (and does not) play a meaningful role in statistical inference. Our examples illustrate that models involving marginalization to non-linear, not one-to-one functions of multiple parameters play a key role in more extreme manifestations of false confidence. 

\vspace{.06in}

\vspace{.06in}
\noindent
{\em Corresponding Author}: Jonathan Williams, \verb5jpwill@live.unc.edu5
\end{abstract}
\end{center}

\pagestyle{plain}
\setcounter{page}{1}

\section{Introduction}

In a recent paper, \cite*{balch2017} presents the phenomenon of ``false confidence'' associated with Bayesian posterior uncertainty.  The authors come about the concept of false confidence from an alarming application to satellite collision risk analysis when estimating the posterior probability of the event that two satellites will collide.  They found that increased measurement error of satellite trajectory data leads to decreased posterior probability of satellites colliding. Essentially, as more noise is introduced into trajectory measurements we become less certain about satellite trajectories, and thus the probability of two satellites colliding decreases.  However, since a posterior probability is an additive belief function (probabilities of mutually exclusive and collectively exhaustive sets sum to one) the probability of the two satellites not colliding must increase accordingly, making their respective trajectories appear safer.  When taken to the extreme, a large enough measurement error will cause an analyst to be (mistakenly) certain the satellites will not collide.  Conversely, when viewed from a likelihood-based sampling distribution framework, more noise in the trajectory data suggests that the satellite trajectories are less certain and therefore are less likely to collide because of the infinitely large number of possible paths they could each take. This alternative interpretation is not problematic.

More on the specifics and importance of satellite collision risk analysis are provided in \cite{balch2017}. To study the mechanics behind what is happening at a more fundamental level the authors present what they term the ``false confidence theorem'' (FCT). This theorem says that with arbitrarily large (sampling/frequentist) probability, there exists a set which does \textit{not} contain the true parameter value, but which has arbitrarily large posterior probability. Such a phenomenon is unsettling for a practitioner making inference based on a posterior distribution.  Moreover, the authors prove that false confidence effects all types of epistemic uncertainty represented by additive probability measures.  This includes Bayesian posterior probabilities, fiducial probabilities, and probabilities derived from most confidence distributions \citep{balch2017}.    

Our goal is to illustrate the intuition and mechanics of the FCT in simple examples so that we can begin to understand more complicated manifestations of the FCT.  Such insight provides a particularly useful contribution to the literature as the use of Bayesian methods becomes more popular. Our contributions in this paper are the following. 

First, we present a simple example to illustrate the mechanics of the FCT with the statistical problem of estimating the support parameter of the U$(0,\theta)$ distribution. This is an example in which the mathematics for the FCT can be worked out analytically and demonstrates where each piece in the statement of the FCT originates from. In most other situations the mathematics cannot be worked out analytically due to the fact that the typical posterior distribution function does not have a readily understood sampling distribution.  In the Appendix we provide similar results for a one parameter Gaussian model.

Next, we show that the FCT manifests in an even more pronounced way by extending the first example to a two parameter model, i.e., U$(0,\theta_{x})$ and U$(0,\theta_{y})$ with $\theta_{x} \ne \theta_{y}$, and considering the marginal posterior distribution of the parameter $\psi = \theta_{x}\theta_{y}$.  This example alludes to the intuition that false confidence is likely at play in situations in which the Gleser-Hwang theorem applies \citep{Gleser1987}.  Such examples are characterized in the frequentist paradigm by exhibiting infinitely large confidence intervals required to obtain less than 100 percent coverage \citep{Berger1999, Gleser1987}. One such famous problem appears in Fieller's theorem \citep{Fieller1954} which has been discussed as recently as the last two meetings of the {\it Bayesian, Fiducial, and Frequentist Conference} (2017, 2018), and in the forthcoming paper \cite*{Fraser2018}.
  
Finally, we demonstrate that the manifestation of the FCT is immediately apparent in a problem related to Fieller's theorem. We show that in reasonable situations the FCT applies to sets which would be concerning in practice. The contribution of such a striking example of false confidence is worrisome in an era in which Bernstein-von Mises type results are unhesitatingly appealed to even when it may not be appropriate (e.g., certain small sample situations).  Such a phenomenon should be properly understood for the appropriate use of Bayesian methodology in practice. 

Broadly, the axioms of probability laid down by \cite{kolmogoroff1933grundbegriffe} have enabled a rich mathematical theory, however, their suitability for modeling epistemic uncertainty has been met with some discontent, particularly the axiom of additivity \citep{shafer2008non}.  The issue with additivity is that it does not leave room for ignorance (i.e., events are either \textit{true} or \textit{false}) which is a major underpinning of the FCT. Theories of inference which weaken additivity assumptions include inferential models \citep{martin2016} and imprecise probabilities \citep{weich2000, gong2017}. 

The paper is organized as follows.  Section \ref{StatementOfFCT} presents and describes the FCT as given in \cite{balch2017}.  Sections \ref{uniform_one_samp}, \ref{uniform_two_samp}, and \ref{normal_two_samp} present and analyze the illustrative examples, and additional analysis is provided in the Appendix.  The $R$ code to reproduce the numerical results presented in this paper is provided at \verb1https://github.com/idc9/FalseConfidence1.

\section{Main ideas}\label{StatementOfFCT}
This section presents the false confidence theorem from \cite{balch2017}. 
\begin{theorem}[\cite*{balch2017}]\label{FCT}
Consider a countably additive belief function $\mathrm{Bel}_{\varTheta\mid X}$ characterized by an epistemic probability density function $\pi_x(\cdot)$ on $\varOmega_\theta$ (the parameter space), with respect to the Lebesgue measure, satisfying $\sup_{\theta\in\varOmega_\theta}\pi_{\mathbf{x}}(\theta)<\infty$, for $P_{\mathbf{X}\mid \theta}$-almost all $x$.  Then, for any $\theta\in\varOmega_\theta$, any $\alpha\in(0,1)$, and any $p\in(0,1)$, there exists a set $A\subseteq\varOmega_\theta$ with positive Lebesgue measure such that $A \not\ni \theta$, and
\begin{equation}\label{FCT_eq}
P_{\mathbf{X}\mid \theta}\big(\big\{ X \colon \mathrm{Bel}_{\varTheta \mid X}(A) \geq 1-\alpha\big\}\big) \geq p.
\end{equation}
\end{theorem}

\begin{wrapfigure}{R}{.45\textwidth}
\centering
\includegraphics[scale=.35]{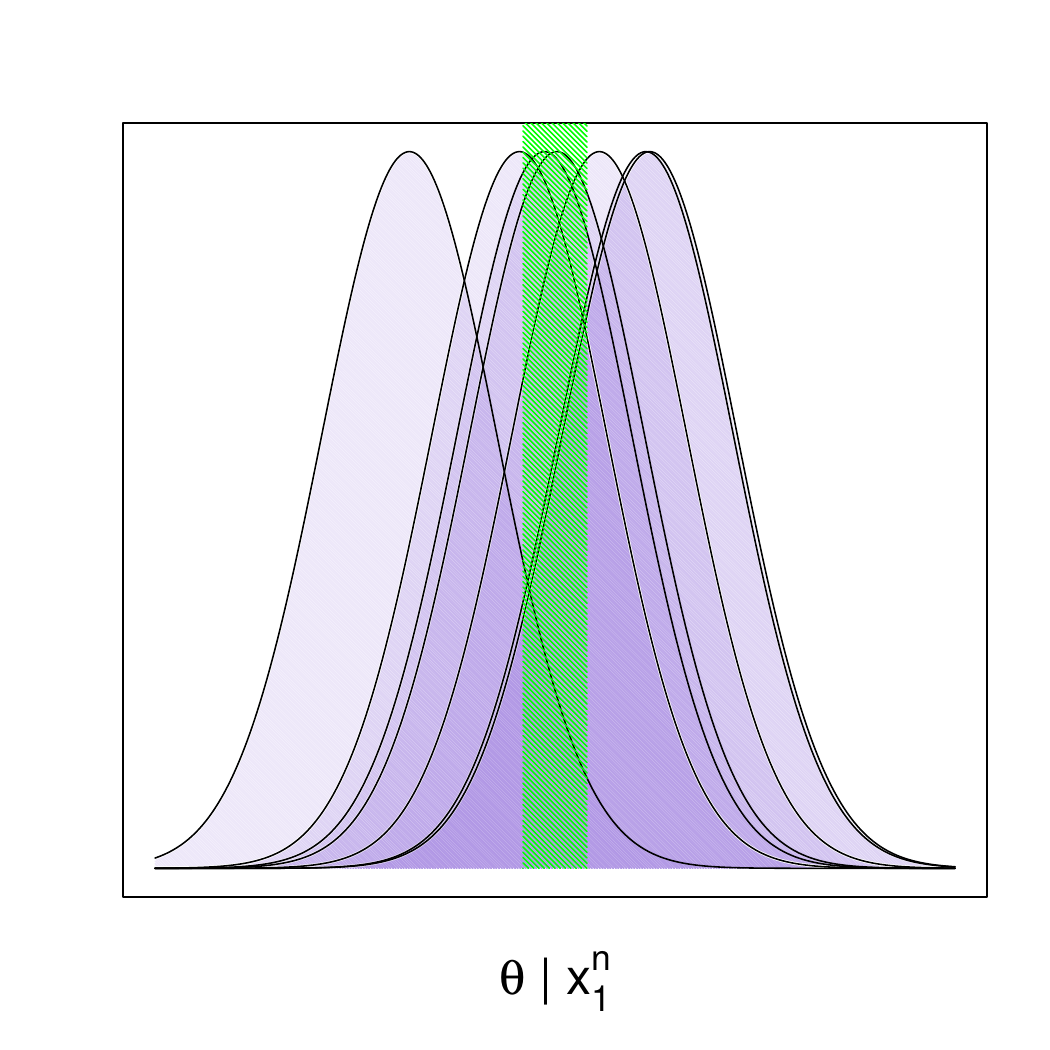}\caption{A sample of realizations from the sampling distribution of the posterior density of the mean, $\theta$, for Gaussian data with known variance and normal prior on $\theta$.  The green shaded region ($A^c$) is an $\varepsilon$-ball around the true parameter value of $\theta$.}\label{sampDistPost}
\vspace{-.3in}
\end{wrapfigure}

While Theorem \ref{FCT} pertains to any form of epistemic probability, for concreteness we will focus on Bayesian posterior probability.  This amounts to considering situations in which
\[\small
\begin{split}
\text{Bel}_{\varTheta\mid X}(A) & = \int_A \pi_{x}(\theta) \ d\theta \\
& = \int_{A}\frac{f_{X\mid\theta}(X)\pi(\theta)}{\int_{\varOmega_\theta} f_{X\mid\vartheta}(X)\pi(\vartheta)~\!\mathrm{d}\vartheta} \ d\theta =: P_{\varTheta\mid X}(A).
\end{split}
\]
To better understand the statement of (\ref{FCT_eq}), Figure \ref{sampDistPost} demonstrates the pieces at play.  The green region represents an example of a particular $A^{c}$ as described in the theorem, and each curve represents a particular realization of the posterior distribution (associated with $P_{\varTheta\mid X}$) over the sampling distribution of the data (associated with $P_{\mathbf{X}\mid \theta}$).

Heuristically speaking, false confidence says that for some set, say $A\subseteq\varOmega_\theta$, which does $not$ contain the true parameter value, the (epistemic) posterior probability $P_{\varTheta\mid X}(A)$ can be made arbitrarily large with arbitrarily large (aleatory) sampling/frequentist probability, i.e., with respect to $P_{X\mid \theta}$. Although the simple existence of such sets $A$ does not immediately raise concerns about statistical inference, for a given situation there may exist practically important sets, such as in the satellite collision risk analysis example of \cite{balch2017}. Note that these sets $A$ may be particularly concerning for finite sample sizes.

The proof given in \cite{balch2017} of the false confidence theorem relies on constructing a neighborhood around the true parameter value.  Accordingly, we investigate further the properties of such sets which satisfy Theorem \ref{FCT} in a few simple and illustrative examples.

\section{Uniform with Jeffreys' prior}\label{uniform_one_samp}

Here we investigate the FCT for uniformly distributed data where the goal is to estimate the support of the distribution.  The motivation for considering this example is that it is simple enough that all of the mathematics can be worked out analytically.  Let $X_{1}, \dots, X_{n}$ be a random sample from the U$(0,\theta)$ distribution where $\theta$ is an unknown parameter. Using the Jeffreys' prior, $\pi(\theta) = 1/\theta$, the posterior will be $\theta \mid X_{1}^{n} \sim \text{Pareto}(n, X_{(n)})$ where $X_{(n)}$ is the maximum of the observed data (see \cite{robert2007bayesian}).

Suppose the true value of $\theta$ is $\theta_0$ and fix $\alpha, p \in (0,1)$.  Then by the proof of Theorem \ref{FCT} (see \cite{balch2017}) there exists $\varepsilon > 0$ such that
\begin{equation}\label{eq:unif_target}
P_{ X_{1}^{n}\mid\theta_{0} }\left(\big\{X_{1}^{n} \colon P_{\theta\mid X_{1}^{n}}(A_\varepsilon)\geq1-\alpha\big\}\right)\geq p,
\end{equation}
where $A_{\varepsilon} \defeq [\theta_0-\varepsilon, \theta_0+\varepsilon]^{c}$, $P_{\theta\mid X_{1}^{n}}$ is the posterior law of $\theta$ (the additive belief function), and $P_{X_{1}^{n}\mid\theta_{0}}$ is the probability measure associated with the sampling distribution of the data.  Note that in this example the Jefferys' prior is a probability matching prior in the Welch-Peers sense (see \cite{Reid2003}); in particular, the interval $C_{x} := (-\infty, X^{(n)}\alpha^{-\frac{1}{n}})$ is such that $P_{\theta\mid X_{1}^{n}}(C_{x}) = 1-\alpha = P_{ X_{1}^{n}\mid\theta_{0} }(X^{(n)}\alpha^{-\frac{1}{n}} \ge \theta_{0})$.  Since the probability matching prior property in one-dimensions pertains to intervals, this fact provides further justification for considering the Jeffreys' prior for analyzing sets of the form $A_{\varepsilon}$.

To compute the left side of (\ref{eq:unif_target}), first re-express as
%P_{ X_{1}^{n}\mid\theta_{0} }\left(\big\{X_{1}^{n} \colon P_{\theta\mid X_{1}^{n}}\big([\theta_0-\varepsilon,\theta_0+\varepsilon]\big)\leq \alpha\big\}\right) & 
\[\small
\begin{split}
& P_{ X_{1}^{n}\mid\theta_{0} }\left( F_{\theta\mid X_{1}^{n}}\left(\theta_{0}+\varepsilon\right) - F_{\theta\mid X_{1}^{n}}\left(\theta_{0}-\varepsilon\right) \leq \alpha \right) \\
&  \hspace{.5in} = P_{ X_{1}^{n}\mid\theta_{0} }\left( 1 - \left(\frac{X_{(n)}}{\theta_{0}+\varepsilon}\right)^{n} - \left[1 - \left(\frac{X_{(n)}}{\theta_{0}-\varepsilon}\right)^{n}\right] \mathbf{1}\{X_{(n)} \le \theta_{0}-\varepsilon\} \leq \alpha \right) \\
&  \hspace{.5in} = P_{ X_{1}^{n}\mid\theta_{0} }\left( \left(\frac{X_{(n)}}{\theta_{0}-\varepsilon}\right)^{n} - \left(\frac{X_{(n)}}{\theta_{0}+\varepsilon}\right)^{n} \leq \alpha \right)\cdot P_{X_{1}^{n}\mid\theta_{0}}(X_{(n)} \le \theta_{0}-\varepsilon) \\
& \hspace{2.5in} +  P_{ X_{1}^{n}\mid\theta_{0} }\left( 1 - \left(\frac{X_{(n)}}{\theta_{0}+\varepsilon}\right)^{n} \leq \alpha \right)\cdot P_{X_{1}^{n}\mid\theta_{0}}(X_{(n)} > \theta_{0}-\varepsilon) \\
&  \hspace{.5in} = P_{ X_{1}^{n}\mid\theta_{0} }\left( X_{(n)} \leq \alpha^{\frac{1}{n}}\left(\frac{1}{(\theta_{0}-\varepsilon)^{n}} - \frac{1}{(\theta_{0}+\varepsilon)^{n}}\right)^{-\frac{1}{n}} \right)\cdot \left(\frac{\theta_{0}-\varepsilon}{\theta_{0}}\right)^{n} \\
& \hspace{2.5in} + P_{ X_{1}^{n}\mid\theta_{0} }\left( X_{(n)} \ge (1 - \alpha)^{\frac{1}{n}} (\theta_{0}+\varepsilon) \right)\cdot \left[1- \left(\frac{\theta_{0}-\varepsilon}{\theta_{0}}\right)^{n}\right]. \\
\end{split}
\]
The second equality comes from the fact that the CDF of the $\text{Pareto}(k, m)$ distribution is given by $F(x) = \big(1 - \big( \frac{m}{x} \big)^k\big) \mathbf{1}\{x \ge m\}$. The third equality comes from considering the two cases of the indicator function, and the final equality comes from solving for $X_{(n)}$.  

\begin{figure}[H]
\centering
\includegraphics[trim=0 2.25in 0 2.25in,clip,scale=.55]{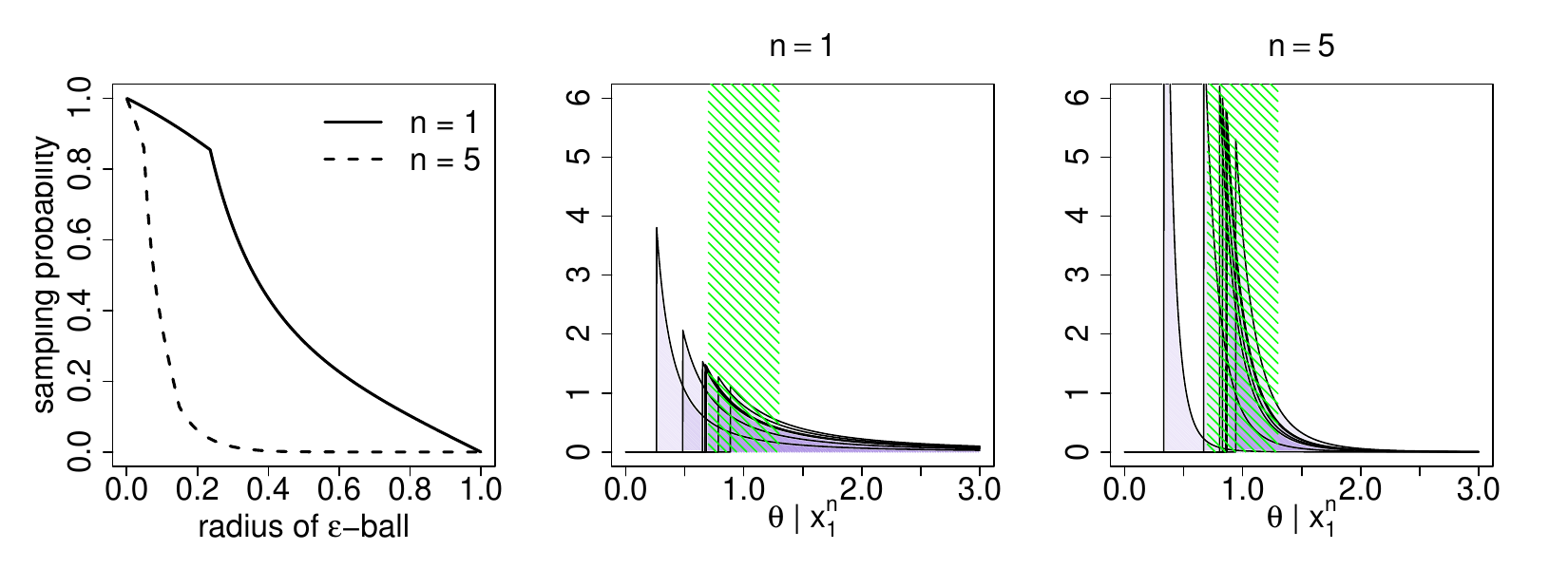}
\caption{The leftmost panel is a plot of the sampling probability, $p$, as a function of $\varepsilon$, as given by equation (\ref{eq:unif_closed_form}), for $\alpha = .5$.  The center and rightmost panels are randomly observed realizations of the posterior density of $\theta$, with a .3-ball around $\theta_{0}$ represented by the shaded green regions.  In all panels, the true parameter value is set at $\theta_{0} = 1$.}\label{one_samp_unif}
\end{figure}

Observe that $\frac{X_{(n)}}{\theta_0} \sim \text{Beta}(n, 1)$ (i.e., maximum order statistic of a U$(0, 1)$ random sample) which gives $P(X_{(n)} \le x) = \big(\frac{x}{\theta_0}\big)^n$. Accordingly,
\begin{equation}\label{eq:unif_closed_form}\small
\begin{split}
& P_{ X_{1}^{n}\mid\theta_{0} }\left(\big\{X_{1}^{n} \colon P_{\theta\mid X_{1}^{n}}\big([\theta_0-\varepsilon,\theta_0+\varepsilon]\big)\leq \alpha\big\}\right) \\
& \hspace{.5in} = \min\left\{ 1, \alpha\left[\left(\frac{\theta_{0}}{\theta_{0}-\varepsilon}\right)^{n} - \left(\frac{\theta_{0}}{\theta_{0}+\varepsilon}\right)^{n}\right]^{-1} \right\} \cdot \left(\frac{\theta_{0}-\varepsilon}{\theta_{0}}\right)^{n} \\
& \hspace{1in} +  \left(1 - (1 - \alpha) \left(\frac{\theta_{0}+\varepsilon}{\theta_{0}}\right)^{n} \right) \mathbf{1}\Big\{\varepsilon \le \theta_{0}\big((1 - \alpha)^{-\frac{1}{n}} - 1\big)\Big\}\cdot \left[1- \left(\frac{\theta_{0}-\varepsilon}{\theta_{0}}\right)^{n}\right]. \\
\end{split}
\end{equation}
Setting the right side of equation (\ref{eq:unif_closed_form}) equal to $p$ gives $p$ as a function of the $\alpha$, $n$, and $\varepsilon$ which satisfy the false confidence theorem. Specifically, we want to know if $\varepsilon$ can be large enough to have a practically meaningful or harmful effect for statistical inference on $\theta_{0}$.  The relationship between $\varepsilon$ and $p$, for $\alpha = .5$, is plotted in Figure \ref{one_samp_unif}.

The leftmost panel in Figure \ref{one_samp_unif} shows, for $\alpha = .5$, the sampling probability (i.e., $p$) that the posterior probability of $A_{\varepsilon}^{c} = [\theta_0 - \varepsilon, \theta_0 + \varepsilon]$ is less than $\alpha$, for $\varepsilon$-balls of various radii.  For example, with $n = 1$ the posterior probability of $A_{\varepsilon}^{c}$ (which contains the true parameter value) will not exceed $.5$ for $\varepsilon \le .3$, for more than 80 percent of realized data sets.  This has the interpretation that the Bayesian test of ``accept $A_{\varepsilon}^{c}$'' if and only if $P_{\theta\mid X_{1}^{n}}(A_{\varepsilon}^{c}) > .5$ would be wrong more than 80 percent of the time.

Displayed on the next two panels of the figure are a few randomly observed realizations of the posterior density of $\theta$, with a .3-ball around $\theta_{0}$ represented by the shaded green regions.  The realizations of the posterior density are typically concentrated around the true value, $\theta_{0} = 1$.  The next section demonstrates how to extend this example into a situation even more amenable to false confidence.

\begin{remark}
This uniform example is one of the few simple examples where we can analytically work out the FCT in a straightforward manner. For example, for interval sets, equation (\ref{eq:unif_target}) shows the posterior CDF needs an analytic sampling distribution.
\end{remark}

\section{Marginal posterior from two uniform distributions}\label{uniform_two_samp}

Assume $X_{1}, \dots, X_{n} \overset{\text{iid}}{\sim} \text{U}(0,\theta_{x})$, and independently $Y_{1}, \dots, Y_{m} \overset{\text{iid}}{\sim} \text{U}(0,\theta_{y})$.  Using the Jeffreys' prior, gives $\theta_{x} \mid X_{1}^{n} \sim \text{Pareto}(n, X_{(n)})$ and $\theta_{y} \mid Y_{1}^{m} \sim \text{Pareto}(m, Y_{(m)})$.  Further, define the nonlinear functional $\psi = \theta_{x}\theta_{y}$, and derive the posterior distribution of $\Psi$ as follows.  By independence,
\[ 
\begin{split}
P_{\Psi \mid X_{1}^{n},Y_{1}^{m}}(\Psi \le \psi) & = \int_{Y_{(m)}}^{\infty} P_{\theta_{x} \mid X_{1}^{n}}\Big(\theta_{x} \le \frac{\psi}{\theta_{y}}\Big) \frac{m Y_{(m)}^{m}}{\theta_{y}^{m+1}} \ d\theta_{y} \\
& = \int_{Y_{(m)}}^{\infty} \Big[1 - \Big(\frac{ X_{(n)} \theta_{y} }{ \psi }\Big)^{n} \Big]\mathbf{1}\Big\{\frac{\psi}{\theta_{y}} \ge X_{(n)}\Big\} \frac{m Y_{(m)}^{m}}{\theta_{y}^{m+1}} \ d\theta_{y}, \\
\end{split}
\]
where the last expression results from the form of the Pareto CDF.  If $n \ne m$, then this equation simplifies to
\[
P_{\Psi \mid X_{1}^{n},Y_{1}^{m}}(\Psi \le \psi) = 1 + \Big(\frac{m}{n - m}\Big) (X_{(n)}Y_{(m)})^{n} \psi^{-n} - \Big(\frac{n}{n - m}\Big)(X_{(n)}Y_{(m)})^{m}\psi^{-m},
\]
and if $n = m$, then the distribution function has the form
\[
P_{\Psi \mid X_{1}^{n},Y_{1}^{m}}(\Psi \le \psi) = 1 - \left[1 + n\log\left(\frac{\psi}{X_{(n)}Y_{(n)}}\right)\right] \left(\frac{X_{(n)}Y_{(n)}}{\psi}\right)^{n}.
\]
In both cases, the support of $\Psi$ is $(X_{(n)}Y_{(m)}, \infty)$.  

\begin{figure}[H]
\centering
\includegraphics[trim=0 2.25in 0 2.25in,clip,scale=.55]{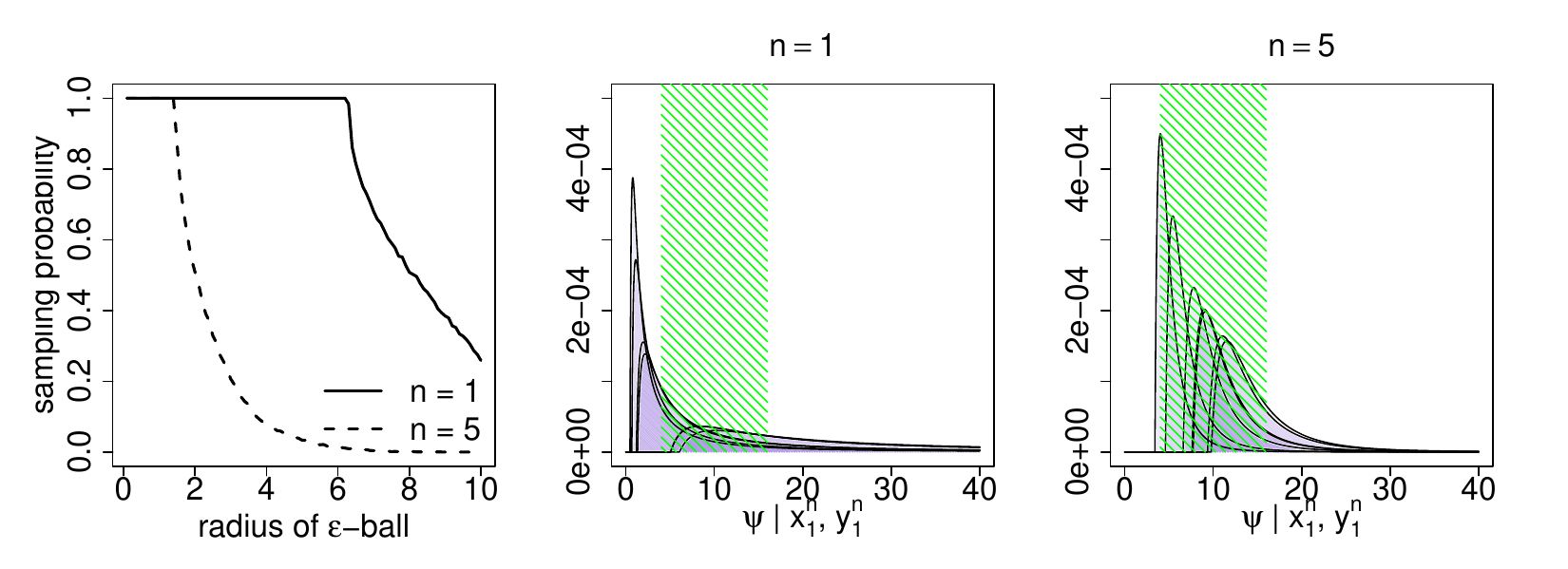}
\caption{The leftmost panel is a plot of the estimated sampling probability, $\widehat{p}_{k}$, as a function of $\varepsilon$, as given by equation (\ref{estimated_p_unif}), for $\alpha = .5$.  The center and rightmost panels are randomly observed realizations of the posterior density of $\Psi$, with a 6-ball around $\psi_{0}$ represented by the shaded green regions.  In all panels, the true parameter value is set at $\psi_{0} = 10$.}\label{two_samp_unif}
\end{figure}

For simplicity, attention will be restricted to the $n = m$ case.  This analytic marginal posterior distribution function makes it simple to estimate $p := P_{X_{1}^{n}, Y_{1}^{n} \mid \psi_{0}}\Big(\big\{X_{1}^{n}, Y_{1}^{n} \colon P_{\psi \mid X_{1}^{n}, Y_{1}^{n}}(A_{\varepsilon}^{c})\leq \alpha \big\}\Big)$, for $A_{\varepsilon}^{c} = [\psi_0 - \varepsilon, \psi_0 + \varepsilon]$ and various values of $\varepsilon$, by simulating data sets and computing the empirical mean, i.e.,
\begin{equation}\label{estimated_p_unif}
\widehat{p}_{k} = \frac{  \# \big\{X_{1}^{n}, Y_{1}^{n} \colon P_{\psi \mid X_{1}^{n}, Y_{1}^{n}}(A_{\varepsilon}^{c})\leq \alpha \big\}  }{ k },
\end{equation}
where $k$ is the number of simulated data set pairs $\{X_{1}^{n}, Y_{1}^{n}\}$.  This is done in Figure \ref{two_samp_unif} for generated data sets.  The true values are set at $\theta_{x}^{0} = 10$ and $\theta_{y}^{0} = 1$ which gives $\psi_{0} = 10$.  Also displayed are a few realizations of the posterior density to illustrate where things go wrong.

From Figure \ref{two_samp_unif} it becomes clear how the FCT manifests.  For $n = 1$, the $\varepsilon$-ball around $\psi_{0} = 10$ with diameter even larger than 12 has posterior probability not exceeding $\alpha = .5$, with sampling probability, $p$, essentially equal to 1.  As in the previous section, this has the interpretation that the Bayesian test of ``accept $A_{\varepsilon}^{c}$'' if and only if $P_{\theta\mid X_{1}^{n}}(A_{\varepsilon}^{c}) > .5$ would essentially always be wrong.  Furthermore, in this case the Bayesian test would fail for an interval (containing the true parameter value) which has length longer than the magnitude of the true parameter value. 

Although this is a toy example being used for pedagogical purposes, it is nonetheless alarming.  One would hope that the small sample size of $n = 1$, while resulting in less posterior certainty about the location of the true parameter value, would be accompanied by more sampling variability/uncertainty.  Rather Figure \ref{two_samp_unif} implies the interpretation that we are $more$ certain about an answer which is in fact false.  The center and rightmost panels of Figure \ref{two_samp_unif} illuminate part of what is happening behind the scene; the posterior densities are typically diffuse around $\psi_{0}$.  The next section presents a more extreme instance of this phenomenon.

\section{Marginal posterior from two Gaussian distributions}\label{normal_two_samp}

Assume $X_{1}, \dots, X_{n} \overset{\text{iid}}{\sim} \text{N}(\theta_{x},\sigma^{2})$, and independently $Y_{1}, \dots, Y_{n} \overset{\text{iid}}{\sim} \text{N}(\theta_{y},\sigma^{2})$.  Suppose also that $\sigma$ is known.  Using independent Jeffreys' priors, gives $\theta_{x} \mid X_{1}^{n} \sim \text{N}(\bar{X}_{n}, \sigma^{2}n^{-1})$ and $\theta_{y} \mid Y_{1}^{n} \sim \text{N}(\bar{Y}_{n}, \sigma^{2}n^{-1})$.  In this context, the nonlinear functional $\psi = \frac{\theta_{x}}{\theta_{y}}$ is related to the classical Fieller's theorem in which infinite confidence intervals are required to attain frequentist coverage \citep{Fieller1954, Gleser1987, Berger1999}.  

The posterior density function for $\psi$ can be derived by transforming the two-dimensional posterior of $(\theta_{x},\theta_{y})$ into the space of $(\psi,\gamma) = (\frac{\theta_{x}}{\theta_{y}},\theta_{y}) =: g(\theta_{x},\theta_{y})$ and then computing the marginal distribution of $\psi$.  Observe that $g^{-1}(\psi,\gamma) = (\psi\gamma,\gamma)$ which gives the Jacobian for the transformation,
$$
J(\psi,\gamma) = \det
\begin{pmatrix}
\gamma & \psi \\
0 & 1 \\
\end{pmatrix} = \gamma.
$$
Then the joint posterior density has the form
\[
\pi_{\psi,\gamma \mid X_{1}^{n}, Y_{1}^{n}}(\psi,\gamma) = \pi_{\theta_{x} \mid X_{1}^{n}}(\psi\gamma) \cdot \pi_{\theta_{y} \mid Y_{1}^{n}}(\gamma) \cdot |\gamma| \cdot \mathbf{1}\{\gamma\ne0\}.
\]
Recalling the forms of the posterior densities for $\theta_{x}$ and $\theta_{y}$, and integrating over $\gamma$ gives
\begin{equation}\label{posterior_dens}
\begin{split}
\pi_{\psi \mid X_{1}^{n}, Y_{1}^{n}}(\psi) & = \int \pi_{\psi,\gamma \mid X_{1}^{n}, Y_{1}^{n}}(\psi,\gamma) \ d\gamma \\
& = \left(\frac{n}{2\pi\sigma^{2}(1 + \psi^{2})}\right)^{\frac{1}{2}} \exp\Big\{\frac{n}{2\sigma^{2}}\left[\frac{(\psi\bar{X}_{n} + \bar{Y}_{n})^{2}}{1 + \psi^{2}} - \bar{X}_{n}^{2} - \bar{Y}_{n}^{2}\right]\Big\} \cdot E_{\gamma | \psi}(|\gamma|), \\
\end{split}
\end{equation}
where the expectation is taken over $\gamma \mid \psi \sim \text{N}\Big(\frac{\psi\bar{X}_{n} + \bar{Y}_{n}}{1 + \psi^{2}}, \frac{\sigma^{2}}{n(1 + \psi^{2})}\Big)$.  

\begin{figure}[H]
\centering
\includegraphics[trim=0 2.25in 0 2.25in,clip,scale=.55]{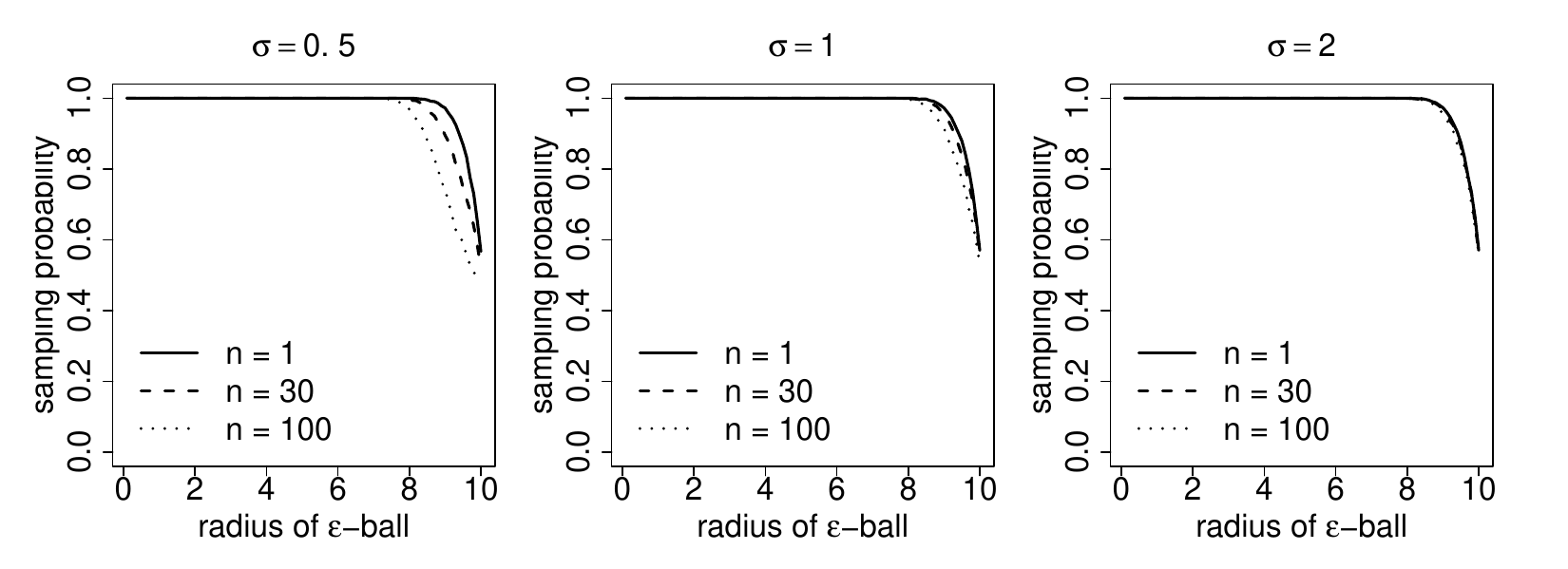}
\caption{Each panel is a plot of the estimated sampling probability of $p$, as a function of $\varepsilon$, using the posterior density equation (\ref{posterior_dens}), and setting $\alpha = .5$.  The true parameter value is $\psi_{0} = 10$.}\label{prob_norm_means_ratio_large}
\end{figure}

\begin{figure}[H]
\centering
\includegraphics[trim=0 2.25in 0 2.25in,clip,scale=.55]{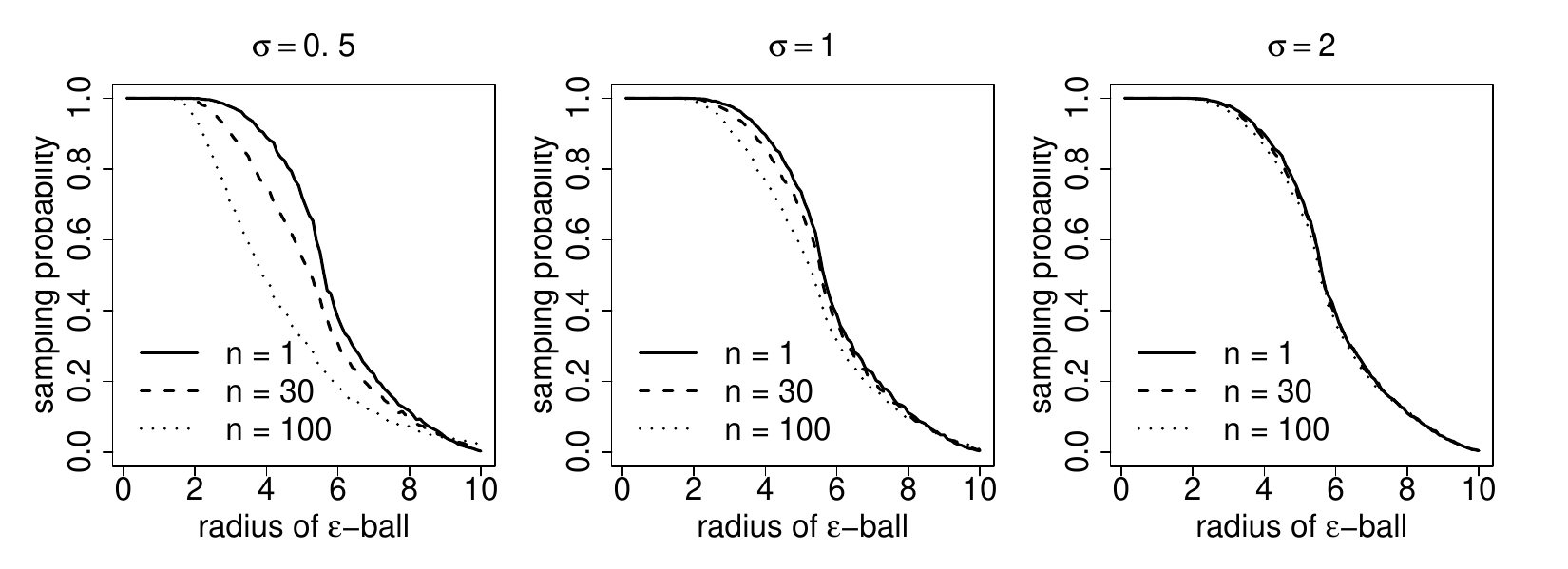}
\caption{Each panel is a plot of the estimated sampling probability of $p$, as a function of $\varepsilon$, using the posterior density equation (\ref{posterior_dens}), and setting $\alpha = .05$.  The true parameter value is $\psi_{0} = 10$.}\label{prob_norm_means_ratio_small}
\end{figure}

\begin{figure}[H]
\centering
\includegraphics[trim=0 2.25in 0 2.25in,clip,scale=.55]{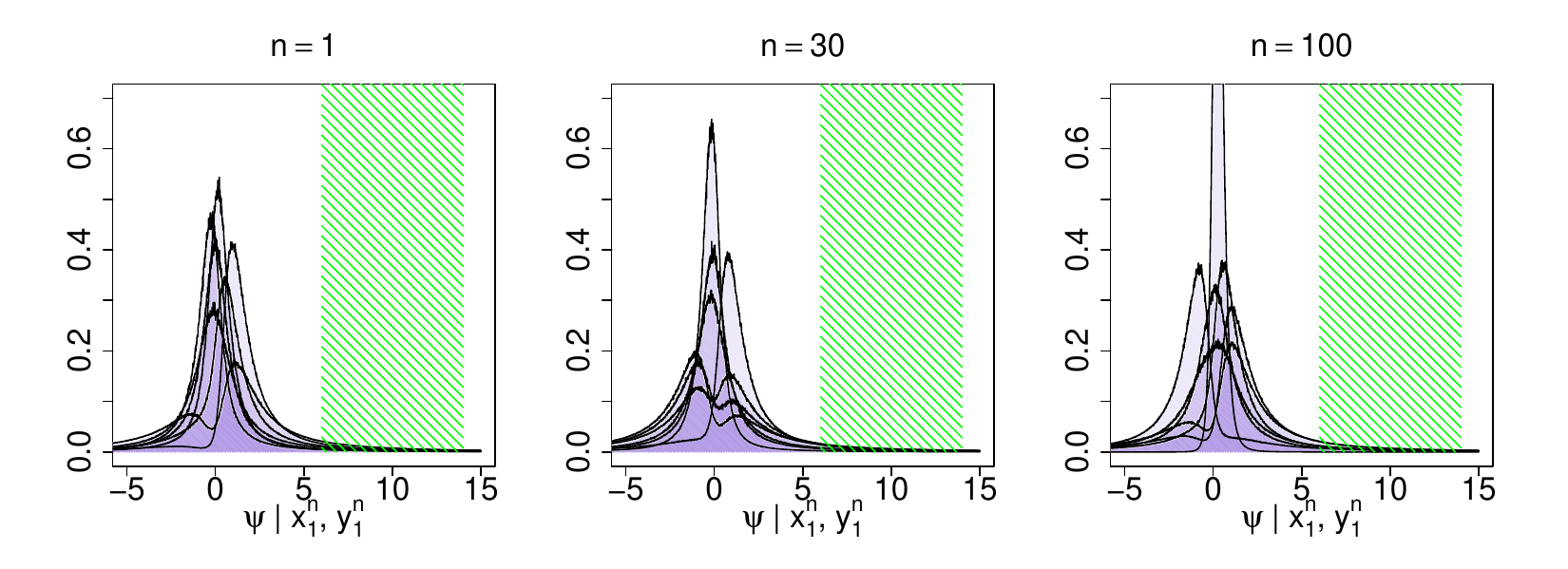}
\caption{Each panel exhibits randomly observed realizations of the posterior density of $\psi$, equation (\ref{posterior_dens}), with a 4-ball around $\psi_{0} = 10$ represented by the shaded green regions.}\label{dens_norm_means_ratio}
\end{figure}

This marginal posterior is easily estimable, and $p := P_{X_{1}^{n}, Y_{1}^{n} \mid \psi_{0}}\big(\big\{X_{1}^{n}, Y_{1}^{n} \colon P_{\psi \mid X_{1}^{n}, Y_{1}^{n}}(A_{\varepsilon}^{c})\leq \alpha \big\}\big)$, for $A_{\varepsilon}^{c} = [\psi_0 - \varepsilon, \psi_0 + \varepsilon]$ and various values of $\varepsilon$, can be estimated with an approximating Riemann sum using equation (\ref{posterior_dens}).  The estimated $p$ as a function of $\varepsilon$ is displayed in Figures \ref{prob_norm_means_ratio_large} and \ref{prob_norm_means_ratio_small} for $\alpha = .5$ and $\alpha = .05$, respectfully, and for various noise levels, $\sigma$.  The true mean values are set at $\theta_{x}^{0} = .1$ and $\theta_{y}^{0} = .01$ which gives $\psi_{0} = 10$.  Displayed in Figure \ref{dens_norm_means_ratio} are a few random realizations of the posterior densities from (\ref{posterior_dens}), for various sample sizes, $n$, with $\sigma = 1$, to illustrate part of where things go wrong.

Remarkably, for almost all values of $n$ and $\sigma$ considered in Figure \ref{prob_norm_means_ratio_large} the Bayesian test of ``accept $A_{\varepsilon}^{c}$'' if and only if $P_{\theta\mid X_{1}^{n}}(A_{\varepsilon}^{c}) > .5$ would fail for $\varepsilon$ as large as 8.  Even considering the extreme choice of $\alpha = .05$ as in Figure \ref{prob_norm_means_ratio_small}, the sampling probability, $p$, exceeds 80 percent chance (in the case of $\sigma = 1$) that $P_{\theta\mid X_{1}^{n}}(A_{\varepsilon}^{c}) \le .05$ for $\varepsilon$ as large as 4, with $n = 100$.

A further illustration of what is happening is once again provided with random realizations of the marginal posterior densities presented in Figure \ref{dens_norm_means_ratio}.  For this problem they heavily concentrate away from the true value $\psi_{0} = 10$.  Consequentially, any inference on the true value of $\psi$ is sure to be misleading, and hence this situation is an extreme example of the manifestation of false confidence in a well-studied classical problem. Similar results hold for the manifestation of false confidence in other non-linear marginalization examples, e.g., the coefficient of variation which is discussed in the Appendix.

\section{Concluding remarks and future work}

There is currently little theoretical understanding of the phenomenon of false confidence or of when it plays a significant role in statistical analysis. We demonstrate ramifications of false confidence in standard, single parameter models as well as models involving the marginalization of multiple parameters. Our examples illustrate that models involving the marginalization to non-linear, not one-to-one functions of multiple parameters seem to play a key role in more extreme manifestations of false confidence.  In future work we seek to gain an understanding of why the FCT is problematic in these situations.

\section{Acknowledgments}

The authors are grateful to Ryan Martin, Jan Hannig, and Samopriya Basu for many helpful comments, engaging conversations, and encouragement.

\appendix
\section{Gaussian with Gaussian prior}

\begin{figure}[H]
\centering\includegraphics[scale=.6]{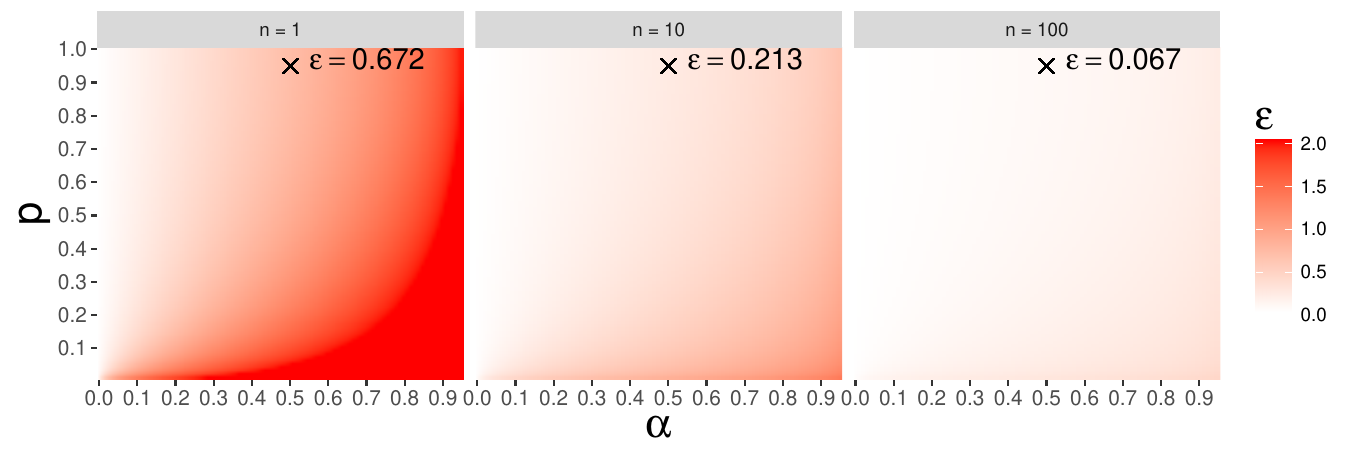}
\caption{Contour plots of $\varepsilon$ as a function of $\alpha$ and $p$ for three different values of $n$ when $\theta_0=1$ and $\sigma^2=1$. The value of $\varepsilon$ for $\alpha=0.5$ and $p=0.95$ is marked with an X.}
\label{fig:normal_contour}
% \vspace{-.2 in}
\end{figure}

Here we provide additional analysis to investigate the FCT for normally distributed data where the goal is to estimate the population mean. Let $X_1,\dots,X_n$ be a random sample from N$(\theta, \sigma^2)$, where $\sigma^2$ is known, but $\theta$ is not and is the object of inference. Consider a prior distribution of $\theta \sim \mathrm{N}(\mu,\tau^2)$.  

%Thus, the likelihood is given by
%\[
%l(\theta\mid X_{1}^{n}) = \left(2\pi\sigma_{0}^{2}\right)^{-\frac{n}{2}}\exp\left(-\frac{1}{2\sigma_0^2}\sum_{j=1}^n(x_j - \theta_0)^2\right),
%\]
%and the prior density is
%\[
%\pi(\theta) = \frac{1}{\sqrt{2\pi\tau^2}}\exp\left(-\frac{(\theta - \mu)^2}{2\tau^{2}}\right).
%\]
Then the posterior distribution is $\theta \mid X_{1}^{n} \sim \mathrm{N}\left(\mu_{n}, \tau_{n}\right)$ where
$\mu_{n} \defeq \left(\frac{\mu}{\tau^2}+\frac{n\bar{X}_n}{\sigma^2}\right)\tau_{n}^{2}$, 
$\tau_{n}^{2} \defeq \left(\frac{1}{\tau^2}+\frac{n}{\sigma^2}\right)^{-1}$,
and $\bar{X}_n\defeq n^{-1}\sum_{j=1}^n X_{j}$. See \cite{hoff2009first} for details.

Suppose the true value of $\theta$ is $\theta_0$ and fix $\alpha, p \in (0,1)$.  Proceeding through the analogous steps as in Sections \ref{uniform_one_samp}-\ref{normal_two_samp} (i.e., we compute $\varepsilon, \alpha$, and $p$ such that equation (\ref{eq:unif_target}) holds),
\begin{align*}
P_{\theta\mid X_{1}^{n}}\big([\theta_0-\varepsilon,\theta_0+\varepsilon]\big) & = \int_{\theta_0-\varepsilon}^{\theta_0+\varepsilon}\frac{1}{\sqrt{2\pi \tau_{n}^{2}}}\exp\left(-\frac{1}{2}\left(\frac{\theta - \mu_{n}}{\tau_{n}}\right)^2\right)\mathrm{d}\theta \\
% & = \int_{\frac{\theta_{0}-\mu_{n}}{\tau_{n}} - \frac{\varepsilon}{\tau_{n}}}^{\frac{\theta_{0}-\mu_{n}}{\tau_{n}} + \frac{\varepsilon}{\tau_{n}}}\frac{1}{\sqrt{2\pi}}\exp\left(-\frac{z^2}{2}\right)\mathrm{d}z\\
& = \Phi\left(\frac{\theta_{0}-\mu_{n}}{\tau_{n}} + \frac{\varepsilon}{\tau_{n}}\right) - \Phi\left(\frac{\theta_{0}-\mu_{n}}{\tau_{n}} - \frac{\varepsilon}{\tau_{n}}\right)
\end{align*}
where $\Phi$ is the standard normal distribution function.  Thus, equation (\ref{eq:unif_target}) here is expressed as 
\begin{equation}\label{eq:SimNormNorm}
P_{ X_{1}^{n}\mid\theta_{0} }\left(\Big\{X_{1}^{n} \colon \Phi\left(\frac{\theta_{0}-\mu_{n}}{\tau_{n}} + \frac{\varepsilon}{\tau_{n}}\right) - \Phi\left(\frac{\theta_{0}-\mu_{n}}{\tau_{n}} - \frac{\varepsilon}{\tau_{n}}\right)\leq \alpha\Big\}\right) \ge p.
\end{equation}
\begin{wrapfigure}{R}{.4\textwidth}
\centering\includegraphics[scale=.6]{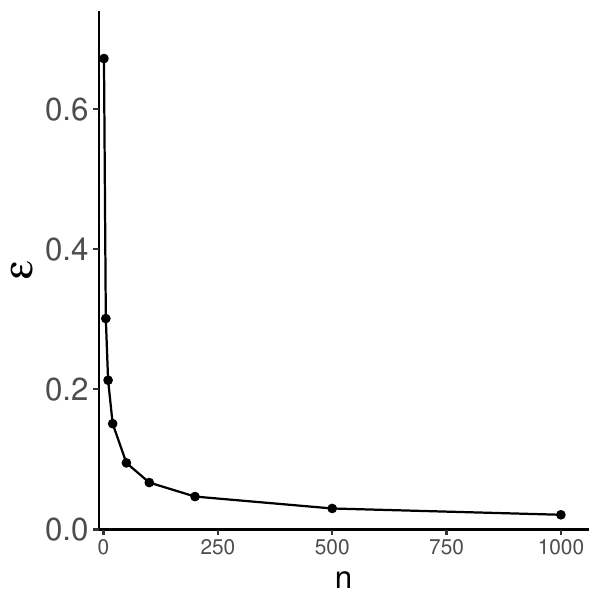}
\caption{Gaussian model. $\varepsilon$ as a function of $n$ where $\alpha$ and $p$ are fixed at $0.5$ and $0.95$, respectively.  The true parameter is $\theta_{0} = 1$.}
\label{fig:normal_normla_n_vs_epsilon}
\vspace{-.1in}
\end{wrapfigure}
Notice that the data appear in (\ref{eq:SimNormNorm}) only through $\mu_{n}$, however, we cannot express $\mu_{n}$ as an analytic function of $\alpha$. If one could do so, then one could define the region of integration to evaluate the outside probability. Hence, a similar analytic expression to equation (\ref{eq:unif_closed_form}) cannot be immediately derived. Therefore, we use Monte Carlo simulation to better understand Equation (\ref{eq:SimNormNorm}).

To make matters concrete, fix $\theta_{0} = 1$, $\sigma^{2} = 1$ (i.e., $X_1^n \overset{\text{iid}}{\sim} \text{N}(1, 1)$), and assign a diffuse prior $\theta \sim \text{N}(0,100)$.  Using Monte-Carlo simulation we compute the value of $\varepsilon$ satisfying equation (\ref{eq:SimNormNorm}) for a range of $\alpha$ and $p$ between 0 and 1, and for the values of $n = 1, 5, 10, 20, 50, 100, 200, 500,$ and $1000$.

Figure \ref{fig:normal_contour} show a contour plot of $\varepsilon$ as a function of $\alpha$ and $p$ for three different values of $n$.  On each of these panels we mark the value of $\varepsilon$ for $\alpha = 0.5$ and $p = 0.95$. This value of $\varepsilon$ has the following meaning: with high sampling probability ($p = 0.95$), a large posterior probability ($1 - \alpha = 0.5$) is assigned to the set $A_{\varepsilon} = [\theta_0 - \varepsilon, \theta_0 + \varepsilon]^c$ which does not contain the true parameter, $\theta_0$. In other words, over repeated sampling of the data, with high probably we will put a lot of belief on values that are at least $\varepsilon$ away from the truth.

The contour plots in Figure \ref{fig:normal_contour} also show that $\varepsilon$ shrinks across the board as $n$ increases. This is made more clear in Figure \ref{fig:normal_normla_n_vs_epsilon} showing $\varepsilon$ as a function of $n$ for fixed $\alpha$ and $p$ (0.5 and 0.95, respectively). For these values of $\alpha$ and $p$, the largest value of $\varepsilon$ is $0.65$ (when $n=3$). 

\section{Coefficient of variation}

\begin{figure}[h]
\centering\includegraphics[scale=.6]{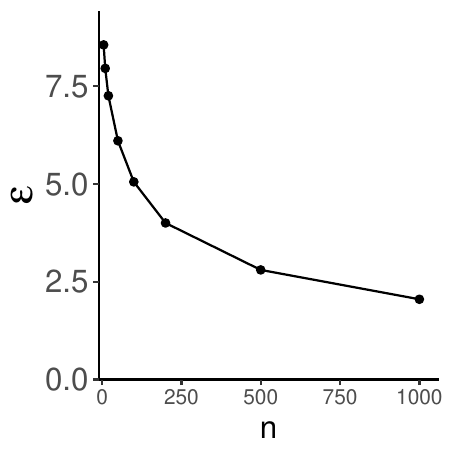}
\caption{Coefficient of variation. $\varepsilon$ as a function of $n$ where $\alpha$ and $p$ are fixed ($0.5$ and $0.9$, respectively). The true parameter is $\psi_0=10$.}
\label{fig:coef_var_n_vs_epsilon}
%\end{wrapfigure}
\end{figure}

Here we consider the coefficient of variation model, and carry out a similar analysis as in the above section.  Let $X_1, \dots X_n \overset{\text{iid}}{\sim} \text{N}(\theta, \sigma^2)$ where both $\theta$ and $\sigma^2$ are unknown. Let $\psi := \frac{\sigma}{\theta}$ be the parameter of interest. The true parameters are taken to be $(\mu_0, \sigma_0) = (1, 10)$ so $\psi_0 = 10$.  Figure \ref{fig:coef_var_n_vs_epsilon} shows $\varepsilon$ as a function of $n$ for $\alpha$ and $p$ fixed ($0.5$ and $0.9$, respectively), and $n=5, 10, 20, 50, 100, 200, 500, 1000$. 

%\begin{wrapfigure}{R}{.4\textwidth}

%%%%%%%%%%%%%%%%%%%%%%%%%%%%%%%%%%%%%%%%%%%%%%%%%%%%%%%%%%%%
\bibliographystyle{agsm}
\bibliography{References}
%%%%%%%%%%%%%%%%%%%%%%%%%%%%%%%%%%%%%%%%%%%%%%%%%%%%%%%%%%%%

\end{document}